\documentclass[10pt, conference, compsocconf]{IEEEtran}

\usepackage[utf8]{inputenc}
\usepackage{amsmath}
\usepackage{amsfonts}
\usepackage{amsthm}

\usepackage{graphicx}
\graphicspath{{Figures/}}
\usepackage{epstopdf}

\usepackage{epsfig}
\usepackage{psfrag}
\usepackage{subfigure}
\usepackage{url}
\usepackage{comment}
\usepackage{afterpage}
\usepackage{cite}
\usepackage{pstricks}
\usepackage{graphics}
\usepackage{soul} % commented by monowar
\usepackage{multirow}

 \usepackage{todonotes} % to do (added by monowar)

\usepackage{graphicx, array, blindtext} % for figure inside table

\usepackage{wrapfig}

\usepackage{algorithm}  
\usepackage{algorithmic}

\usepackage{listings}
\usepackage{psfrag,wrapfig}

\usepackage{enumerate}% http://ctan.org/pkg/enumerate

\newcommand{\eg}{{\it e.g.,}\xspace}
\newcommand{\viz}{{\it viz.,}\xspace}

\newcommand{\etc}{{\it etc.~}}

\usepackage{amsthm}
\theoremstyle{definition}
\usepackage{multirow}

\newcounter{myoptimizationproblemctr}
\newenvironment{myoptimizationproblem}{
   \bigskip\noindent%         create a vertical offset to previous material
   \refstepcounter{myoptimizationproblemctr}% increment the environment's counter
   $(\mathbf{P\themyoptimizationproblemctr})$ %prefix text   
   }{}   % {\par\bigskip}  %          create a vertical offset to following material
\newtheorem{theorem}{Theorem}
\theoremstyle{definition}

\theoremstyle{remark}

\theoremstyle{definition}

\newtheorem{proposition}{Proposition}
\newtheorem{observation}{Observation}

\begin{document}

%\title{OMonitor: A Framework for Opportunistic Monitoring in Hard Real-Time Systems}

%\title{A Framework for Integrating Security Policies in Hard Real-Time Systems}

\title{Exploring Opportunistic Execution for Integrating Security into Legacy Hard Real-Time Systems}
%\title{Securing Hard Real-Time Systems with Opportunistic Execution}

%\title{Title}

\author{\IEEEauthorblockN{Monowar Hasan\IEEEauthorrefmark{1}, Sibin Mohan\IEEEauthorrefmark{1},  Rakesh B. Bobba\IEEEauthorrefmark{2} and Rodolfo Pellizzoni\IEEEauthorrefmark{3}} \IEEEauthorblockA{\IEEEauthorrefmark{1}Dept. of Computer Science, University of Illinois at Urbana-Champaign, Urbana, IL, USA}
\IEEEauthorblockA{\IEEEauthorrefmark{2}School of Electrical Engineering and Computer Science, Oregon State University, Corvallis, OR, USA}
\IEEEauthorblockA{\IEEEauthorrefmark{3}Dept. of Electrical and Computer Engineering, University of Waterloo, Ontario, Canada}
Email: \{\IEEEauthorrefmark{1}mhasan11, \IEEEauthorrefmark{1}sibin\}@illinois.edu,
\IEEEauthorrefmark{2}rakesh.bobba@oregonstate.edu,
\IEEEauthorrefmark{3}rodolfo.pellizzoni@uwaterloo.ca}

\maketitle

% should be commented out for any final version
\thispagestyle{plain}
\pagestyle{plain}

 \begin{abstract}
Due to physical isolation as well as use of proprietary hardware and protocols, traditional real-time systems (RTS) were considered to be invulnerable to security breaches and external attacks. However, this assumption is being challenged by recent attacks that highlight the vulnerabilities in such systems. In this paper, we focus on integrating security mechanisms into RTS (especially {\em legacy} RTS) and provide a {\em metric} to measure the effectiveness of such mechanisms. We combine opportunistic execution with hierarchical scheduling to maintain compatibility with legacy systems while still providing flexibility. The proposed approach is shown to increase the security posture of RTS systems without impacting their temporal constraints.

%security tasks to be compatible with legacy systems and scheduling server based approach to  aim to detect malicious activities by executing security mechanisms as independent sporadic tasks, without violating temporal constrains of the real-time control tasks. The presented mechanism that integrates security mechanisms will increase the difficulty for would-be attackers, and thus improve the overall security posture for hard RTS.
 \end{abstract}

\section{Introduction}

Real-time Systems (RTS) are everywhere. Embedded RTS are used for monitoring and controlling physical systems and processes in varied domains, \eg aircraft including Unmanned Aerial Vehicles (UAVs), submarines, other vehicles (both autonomous as well as manual), critical infrastructures (\eg  power grids and water systems), spacecraft and industrial plants to name but a few. These systems rely on a variety of inputs for their correct operation and have to meet stringent safety and timing requirements. However, until recently, cyber-security considerations were an afterthought in the design of such systems. While fault-tolerance has been a design consideration, traditional fault-tolerance techniques that were designed to counter and survive random or accidental faults are not sufficient to deal with cyber-attacks. Given the increasing cyber-attack risks, it is essential to integrate resilience against such attacks into RTS design and to retrofit existing critical RTS with detection, survival and recovery mechanisms. However, any such mechanisms have to co-exist with the real-time tasks in the system and have to operate without impacting the timing and safety constraints of the control logic. This creates an apparent tension, especially in the case of legacy systems -- between security requirements (\eg having enough cycles for effective detection) as well as the timing and safety requirements. Further, any detection mechanism has to be designed so that an adversary cannot easily evade it. 

The stringent timing constraints in hard real-time systems introduce additional complexities for the implementation of cyber-security mechanisms; \eg the strict deadlines for the completion of periodic hard real-time systems may not allow for additional execution by the security mechanisms. Unlike conventional IT security solutions, it may not be possible to execute the security tasks for arbitrary periods of time. This is because, not only must the security mechanisms work properly but they must also not interfere with the deadlines of real-time tasks.

\begin{table*}[!htb]
\caption{Example of security tasks for intrusion detection in Linux-based RTOS\textsuperscript{*}}
\label{table:rtos}

\centering
\begin{tabular}{p{4.0cm} p{7.2cm} p{0.5cm} m{2.8cm}}
%\toprule
\hline
Task & Function & Criticality & $\quad$ Execution Time \\
%\midrule
\hline
\hline 
Check IDS own binary (IDS\_bin) & Scan (\eg compare hash value) files in the following locations: $\mathtt{/usr/sbin/siggen}$, $\mathtt{/usr/sbin/tripwire}$, $\mathtt{/usr/sbin/twadmin}$, $\mathtt{/usr/sbin/twprint}$  & High & \vspace*{0.05em}\multirow{3}{*}{ 
\includegraphics[scale=0.21]{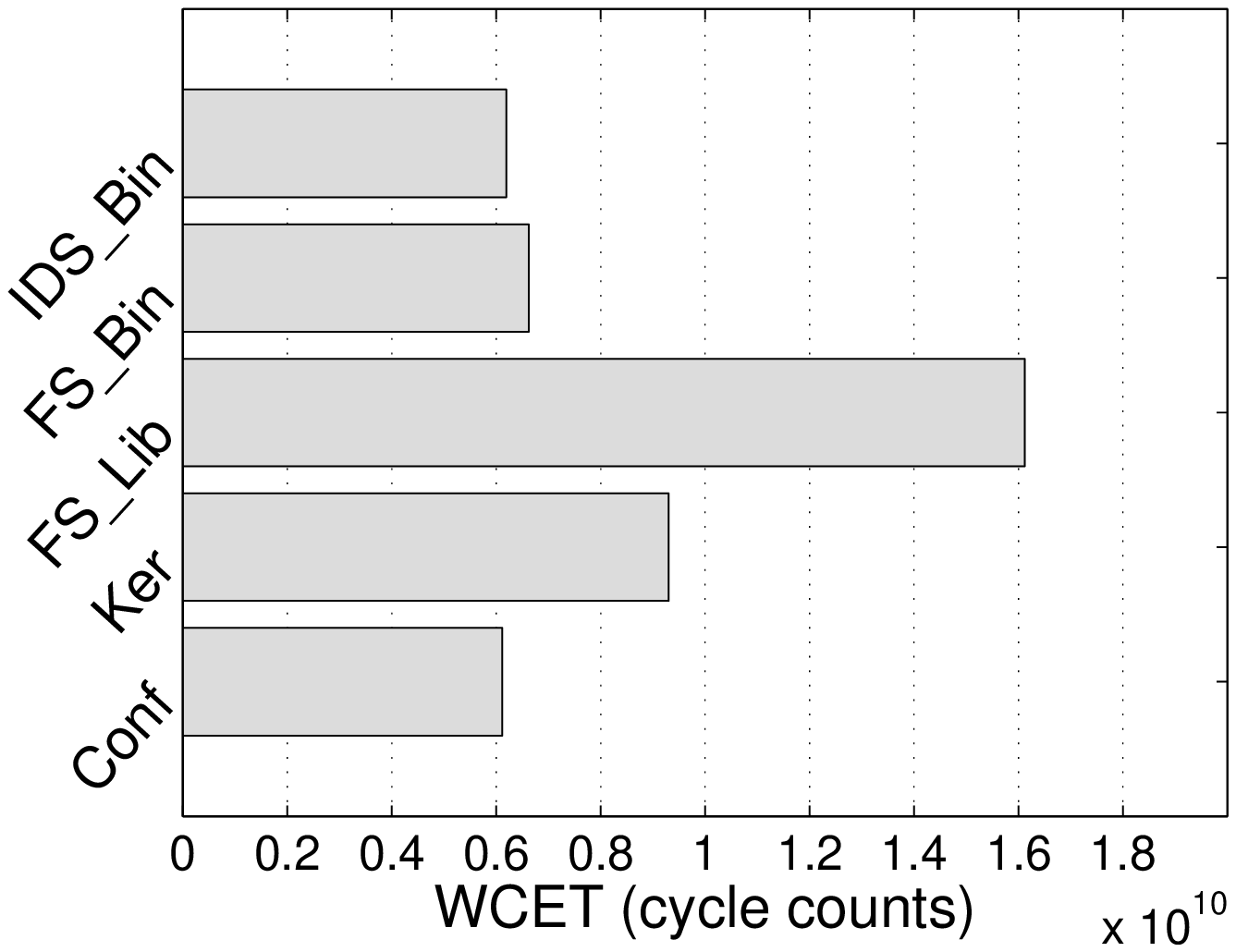}}  \\
Check critical executables (FS\_bin) & Scan file-system binary (\eg $\mathtt{/bin}$, $\mathtt{/sbin}$) & High  \\
Check critical libraries (FS\_lib) & Scan file-system library (\eg $\mathtt{/lib}$)  & High  \\
Check device and kernel (Ker) & Scan peripherals and kernel information in the  $\mathtt{/dev}$ and $\mathtt{/proc}$ directory & High  \\
Check configuration files (Conf) & Scan changes in the configuration files (\eg $\mathtt{/etc}$) & Medium \\

%\midrule
\hline 
\multicolumn{4}{p{15.5cm}}{\textsuperscript{*}\footnotesize{We use Linux kernel version 3.10.32 with real-time patch, \eg Real-Time Application Interface (RTAI) \cite{rtai} version 4.1. The execution times are measured using  read time-stamp counter (RDTSC) \cite{rdtsc} instruction.}}
%\bottomrule
\end{tabular}

\end{table*} 

Our goal is to improve RTS security posture by integrating security mechanisms \textit{without} violating real-time constraints. The security routines could be any intrusion detection and recovery mechanism depending on the system requirements. As an example, let us consider an open source intrusion detection mechanism, Tripwire\footnote{\url{http://www.tripwire.com/}.}, that detects integrity violations in the system. It stores clean system state during initialization and uses it later to detect intrusions by comparing the current system state against the stored clean values. As we show in Table \ref{table:rtos}, the default configuration of Tripwire contains several tasks, \viz  protecting its own binary files, protecting system binary and library files, ensuring kernel and process integrity \etc 

%In practical systems, 
While integrating security mechanisms into a practical system, the following performance criteria need to be considered. 

\begin{enumerate}[\itshape i\normalfont)]
\item \textit{Monitoring Frequency:} In order to provide best protection, the security tasks need to be executed quite often. If the interval between consecutive monitoring events is too large, the adversary may harm the system undetected between two invocations of the security task. On the other hand, if the security tasks are executed very frequently, it may impact the schedulability of the real-time tasks. Herein lies an important trade-off between monitoring frequency and schedulability. 

\item \textit{Responsiveness:} In some circumstances, a security task may need to complete with less interference from higher-priority tasks. As an example, let us consider the scenario where a security breach is suspected. %and the system needs to be examined more thoroughly%to detect for the adversary. %In such cases, in order to make the system secure, the security routines may require to executed with higher priority by pausing some less-critical real-time tasks. 
In such an event the security task may be required to \textit{perform more fine-grained checking instead of waiting for its next sporadic slot}. This may cause some low-priority non-critical real-time tasks to miss their deadlines. However, the scheduling policy needs to ensure that the system remains secure without violating real-time constraints for critical, high-priority, real-time tasks. 

\item \textit{Atomicity:} Depending on the operation, some of the security tasks may need to be executed \textit{without preemption}. For instance, let us consider a security task that scans the process table and has been preempted in the middle of its operation. An adversary may corrupt the process table entry that has already been scanned before the next scheduling point of the security task. When the security tasks are rescheduled, it will start scanning from its last state and may not be able to detect the changes in a timely manner.  
\end{enumerate}

In this paper we focus mainly on the {\it monitoring frequency} criterion\footnote{As the other two imply changing the schedule of real-time tasks which in most cases is not feasible with legacy systems.}. However, we  highlight how our proposed framework can be extended to incorporate {\it responsiveness} and \textit{atomicity} in Section \ref{sec:lim_n_dis}. In particular, we consider incorporating security mechanisms by implementing them as separate \textit{sporadic tasks}. This brings up the challenge of determining the `right' periods for the security tasks. For instance, some critical security routines may be required to execute more frequently than others. However, if the period is too short (\eg the security task repeats too often) then it will use too much of the processor time and eventually lower the overall system utilization. As a result, the security mechanism itself might prove to be a hindrance to the system and reduce the overall functionality or worse, safety. On the other hand, if the period is too long, the security task may not always detect violations since attacks could be launched between two instances of the security task. 

Therefore, we propose a framework that allows the execution of security tasks \textit{opportunistically} with lower-priority than real-time tasks, while keeping the best possible periods for the security tasks that ensure all the tasks in the system remain schedulable. The main contribution of this paper can be summarized as follows:

\begin{itemize}
\item We introduce an extensible framework that allows the security tasks to execute without perturbing scheduling order and timing constraints of the real-time tasks (Section \ref{sec:algorithm}). In doing so, we formulate a constraint optimization problem and solve it using Geometric Programming (GP) approach in polynomial time (Sections \ref{sec:period_adapt}-\ref{sec:server_param_selec}). 

\item We propose a metric to measure the security posture of the system in terms of frequency of periodic execution (Section \ref{sec:period_adapt}).

\item We evaluate the proposed approach for schedulability and security (Section \ref{sec:evaluation}).

\end{itemize}

%We organize rest of the paper as follows. We first present an adversary and system model in Section \ref{sec:sys_model}. Followed by the analytical studies in Section \ref{sec:period_adapt}, \ref{sec:sec_server}, and \ref{sec:server_param_selec}, the overall algorithm is presented in Section \ref{sec:algorithm}. Section \ref{sec:evaluation} presents the experimental results. In Section \ref{sec:lim_n_dis} we discuss the limitations and possible extensions of this work. The related work is surveyed in Section \ref{sec:related_works},  before we conclude the paper in Section \ref{sec:conclusion}. 
Key mathematical notations used in the paper are listed in Table \ref{tab:notations}.

\begin{table}[!htb]
\renewcommand{\arraystretch}{1.1}
\caption{Mathematical Notations}
\label{tab:notations}
\centering
\begin{tabular}{c||p{5.9cm}}
\hline %\\
\bfseries Notation & \bfseries Interpretation\\
\hline\hline
$\Gamma_R$, $\Gamma_S$ & Set of real-time and security tasks, respectively \\
$m$ , $n$ & Number of real-time and security tasks, respectively \\
$C_i$, $D_i$, $T_i$ & Worst-case execution time, deadline, and period of task $\tau_i$, respectively \\
$hp(\tau_i)$ & Set of tasks that has higher priority than $\tau_i$ \\
$T_i^{des}$, $T_i^{max}$ & Desired and maximum allowable period of the security task $\tau_i \in \Gamma_S$, respectively \\
$\omega_i$ & Weighting factor of the security task $\tau_i \in \Gamma_S$\\
$\mathcal{S}(Q, P)$ & Security server with capacity $Q$ and replenishment period $P$ \\
$\tau_{\mathcal{S}}$ & The task represents the security server $\mathcal{S}(Q, P)$ with capacity $Q$ and replenishment period $P$ \\
$UB_{\mathcal{S}(Q, P), \Gamma_S} $ & Utilization bound for the given server parameters $\mathcal{S}(Q, P)$ and security task-set $\Gamma_S$ \\
$\Delta_{\mathcal{S}}$ & Worst case interference to the security server by the real-time tasks \\
$I_i$ & Worst-case workload generated by the $\tau_i$ and $hp(\tau_i)$ from critical instance to deadline of $\tau_i$ \\
$\mathbf{lsbf}_{\mathcal{S}}(t)$  & Linear lower-bound supply function of the security server during time interval $t$ \\
\hline
\end{tabular}
\end{table}

\section{System and Security Model}
\label{sec:sys_model}

\subsection{Real-Time Tasks and Scheduling Model}
\label{subsec:task_model}

We consider a uni-processor system consisting of a set of $m$ fixed-priority\footnote{We assume that task priority assignment follows Rate Monotonic (RM) \cite{Liu_n_Layland1973} algorithm.} 
sporadic real-time tasks $\Gamma_R = \lbrace \tau_1, \tau_2, \cdots , \tau_m \rbrace$. Each real-time task $\tau_i \in \Gamma_R$ is characterized by $(C_i, T_i , D_i)$, where $C_i$ worst-case execution time (WCET), $T_i$ is the minimum inter-arrival time (or period) between successive releases and $D_i$ is the relative deadline.  We assume that priorities are distinct and the tasks have implicit deadline, \eg $D_i = T_i$ for $\forall i \in \Gamma_R$. %The processor utilization of $\tau_i$ is defined as $U_i = \frac{C_i}{T_i}$. Each real-time task releases a (potentially infinite) sequence of jobs. %at each invocation. 
%Since $D_i = T_i$ when the tasks are schedulable, there can be at most one job per task at any time instant. 
For the simplicity of  notation we use the same symbol $\tau_i$ to denote a task's jobs. % for and hence, we use the terms, task and job, interchangeably.

%. Each task τi releases a (potentially infinite) sequence of jobs, with the first job released at any time during the system execution and subsequent jobs released at least Ti (referred to as minimum inter-arrival time) time units apart

Let $hp(\tau_i)$ denote the set of tasks that have higher priority than $\tau_i$. We assume that the real-time task-set $\Gamma_R$ is \textit{schedulable} by a fixed-priority preemptive scheduling. Since the task-set is schedulable, for each task $\tau_i \in \Gamma_R$, the worst-case response time $w_i$ is less than or equal to the deadline $D_i$. Hence, the following inequity is satisfied for all tasks $\tau_i \in \Gamma_R$  \cite{res_time_1, res_time_rts}:
\begin{equation} \label{eq:wcrt}
w_i = C_i + \sum_{\tau_h \in hp(\tau_i)} \left\lceil \frac{r_i^k}{T_h} \right\rceil \cdot C_h \leq D_i
\end{equation}
where $r_i^0 = C_i$ and $w_i = r_i^{k+1} = r_i^k$ for some $k$.  

\subsection{Security Model}
\label{subsec:sec_model}

RTS face threats in various forms depending on the system and the goals of an adversary. For instance, adversaries may eavesdrop on, insert or modify messages exchanged by RTS; they may manipulate the sensor inputs being processed; the adversary could try to modify the control flow of the system; glean sensitive information through side channels \etc The goals of adversaries may range from simply lodging themselves in the system and stealthily learning  sensitive information to actively taking control and manipulating or crashing the system.

Threats to communications are usually dealt with by integrating cryptographic protection mechanisms. From an RTS perspective this increases the WCET of existing real-time tasks and has been studied in literature \cite{xie2007improving, lin2009static}. In contrast, our focus is on threats that can be dealt with by integrating additional security tasks. For example, a sensor measurement correlation task may be added for detecting sensor manipulation or a change detection task may be added to detect changes or intrusions into the system. The addition of such tasks may necessitate changing the schedule of real-time tasks as was the case in \cite{sg1, sg2} where a state cleansing task was added to deal with stealthy adversaries gleaning sensitive information through side channels. In this work we focus on situations where added security tasks are not allowed to impact the schedule of existing real-time tasks as is often the case when integrating security into legacy systems. However, we do discuss the extensiblity of our approach to scenarios when some changes may be allowed.

%RBB: commented out by RBB on May 4th 2016
\begin{comment}
We assume that an adversary may either able to insert a malicious task that respects the real-time guarantees of the system to avoid immediate detection; and/or can be able to compromise one or more existing real-time tasks to launch the attack. Besides, we assume that the scheduler is trustworthy and schedules the security tasks based on the parameters obtained by the proposed framework. The adversary may aim to destabilize the system and/or glean important information either by side-channel or covert-channel attacks. Hence for best security enforcement, we intend to execute security routines quite frequently.

Other than trying to aggressively crash the system, the intruder may utilize side-channels to monitor the system behavior and infer certain degree of system information (\eg hardware/software architecture, user task and thermal profiles etc.). Depending on the actual implementation of the security routines, such side-channel attacks may or may not be detected by our proposed monitoring mechanism. Beside, we limit our study to the host-based intrusion detection mechanisms. For instance, an advanced attacker may capture I/O or network ports and perform network-level attacks and tamper the confidentiality and integrity of the system. Again, capturing those attack behavior is limited by the actual implementation of the security routines itself and hence we consider generic host-based intrusion detection mechanisms as our security policy. 
\end{comment}

\subsection{Security Tasks} \label{subsec:sec_task}

 %With a view to imposing  security in the system, we want to integrate additional $n$ fixed-priority sporadic security tasks. 
Our goal is to ensure security of the system \textit{without} perturbing the scheduling order and  timing constraint (\eg Eq. (\ref{eq:wcrt})) of the real-time tasks. As mentioned earlier, we ensure security of the system by integrating additional security tasks. Let us denote the security tasks by the set $\Gamma_S = \lbrace \tau_1, \tau_2, \cdots , \tau_n \rbrace$. Each security task $\tau_i \in \Gamma_S$ is characterized by $(C_i, T_i^{des}, T_i^{max}, \omega_i)$, where $C_i$ is the WCET, $T_i^{des}$ is the most desired period between successive releases (hence $F_i^{des} = \frac{1}{T_i^{des}}$ is the desired execution frequency of security routine), and $T_i^{max}$ is the maximum allowable period beyond which security checking by $\tau_i$ may not be effective. The parameter $\omega_i > 0$ is a designer given weighting factor, that may reflect the criticality of the security task $\tau_i$. %For instance, 
More critical security tasks would have larger $w_i$. For example, as illustrated in Table \ref{table:rtos}, the default configuration of Tripwire has different criticality levels, \eg  \textit{High} (critical files that are significant points of vulnerability), \textit{Medium} (non-critical files that are of significant security impact) etc.

Any period $T_i$ within the range $T_i^{des} \leq T_i \leq T_i^{max}$ would be acceptable. The actual period $T_i$, however, is not known a priori and our goal is to find the suitable period for the security tasks without violating the real-time constraints. We assume that the security tasks also have implicit deadline, \eg $D_i = T_i$ for $\forall \tau_i \in \Gamma_S$ that implies security tasks should complete before its next period. 

 %for the security tasks

%desired frequency (\eg security task period) of periodic monitoring.

 %the arrivals of tasks are independent. Since each task can be executed the moment it arrives and , and thus two or more tasks may be released at the same time.

\section{Period Adaptation} \label{sec:period_adapt}

A simple approach to ensure security without perturbing real-time scheduling order is to execute security tasks as lower priority than real-time tasks. Hence, the security routines will be executing only during slack times when no other higher-priority real time tasks are running. However, one fundamental problem is to determine \textit{which} security tasks will be running \textit{when} to provide better defense against certain vulnerabilities. As mentioned earlier, actual period of the security tasks is unknown and we need to \textit{adapt} the periods within acceptable ranges to achieve better trade-off between schedulabily and defense against security breaches. In what follows, we formulate the period adaptation problem to ensure security tasks can execute close to their desired frequency. 

%In this section we present the period adaptation problem for the given server parameters (\eg capacity and replenishment period)\footnote{The calculation of  capacity and replenishment period is explained detail in Section \ref{sec:server_param_selec}.}. We first formulate the problem as a non-linear constraint optimization problem and later transform it to a convex geometric program which can be solved very efficiently.

\subsection{Problem Description}

We measure the security of the system by means of the \textit{achievable periodic monitoring}. Let $T_i$ be the period of the security task $\tau_i \in \Gamma_S$ that needs to be determined. Our goal is to minimize the perturbation between the achievable period $T_i$ and the desired period $T_i^{des}$. Hence we define the metric 
\begin{equation}
\eta_i = \frac{T_i^{des}}{T_i}
\end{equation}
that denotes the \textit{tightness} of the frequency of periodic monitoring for the security task $\tau_i$ and bounded by $\frac{T_i^{des}}{T_i^{max}} \leq \eta_i \leq 1$. The period adaptation problem is therefore to maximize the tightness of the achievable periods of all the security tasks without violating the schedulabulity as well as period bound constraints of the security tasks. The formulation of the period adaptation problem is explained in the following. 

\subsection{Formulation as an Optimization Problem}

\subsubsection{Objective Function}

As mentioned earlier, the objective of the period adaptation is to minimize the perturbation (\eg maximize the tightness $\eta_i$) for all the security tasks. Mathematically the objective function can be defined as follows:
\begin{equation} \label{eq:period_obj}
\underset{\mathbf{T}}{\operatorname{max}} ~  \eta %= \underset{\mathbf{T}}{\operatorname{max}} ~  \sum_{\tau_i \in \Gamma_S} \omega_i \frac{T_i^{des}}{T_i}
\end{equation}
where $\eta = \sum\limits_{\tau_i \in \Gamma_S} \omega_i \eta_i = \sum\limits_{\tau_i \in \Gamma_S} \omega_i \frac{T_i^{des}}{T_i}$ is the cumulative weighted tightness and the vector $\mathbf{T} = [ T_1, T_2, \cdots, T_n ]^{\mathsf{T}}$ represents the period of the tasks that need to be determined.

%and the constant $\omega_i > 0$ is a given weighting factor that reflects criticality of the security task $\tau_i$. More critical security tasks would have larger $w_i$. 

\subsubsection{Utilization Bound Constraint}

Utilization of each security task is an important measure since it reveals the amount of processor dedicated to the security task and thus impacts schedulability. 
Since we are executing security tasks in the idle time, the following necessary condition ensures that utilization of the security tasks are within the remaining utilization \cite{Liu_n_Layland1973}: %\todo{need citation here (added -- M)}
\begin{equation} \label{eq:period_ns_const_2}
%\sum_{\tau_i \in \Gamma_S}\frac{C_i}{T_i} \leq \ln 2 - \sum_{\tau_i \in \Gamma_R}\frac{C_i}{T_i}.
\sum_{\tau_i \in \Gamma_S}\frac{C_i}{T_i} \leq (m+n)(2^{\frac{1}{m+n}} - 1) - \sum_{\tau_j \in \Gamma_R}\frac{C_j}{T_j}.
\end{equation}

\subsubsection{Period Bound Constraints}

Since we consider that the integrated system consisting of real-time and security tasks will follow the RM priority order, the following constraint needs to be satisfied
\begin{equation} \label{eq:period_ns_const_rm}
T_i \geq \bar{T} \quad \forall \tau_i \in \Gamma_S
\end{equation}
where $\bar{T} = \underset{\tau_j \in \Gamma_R}{\operatorname{max}} T_j$. Besides, in order to fulfill the restrictions on periodic monitoring, %the period restrictions, 
the following inequality needs to be satisfied for all the security tasks
\begin{equation} \label{eq:period_const_1}
T_i^{des} \leq T_i \leq T_i^{max}  \quad \forall \tau_i \in \Gamma_S.
\end{equation}

Using the objective function in Eq. (\ref{eq:period_obj}), and the set of constraints in Eqs. (\ref{eq:period_ns_const_2}), (\ref{eq:period_ns_const_rm}), and (\ref{eq:period_const_1}), we can formulate the period adaptation problem as the following non-linear constraint optimization problem.

\begin{myoptimizationproblem} \label{opt:period_adapt_ns}
\vspace*{-2.0em}
\begin{subequations}
\begin{align}
\underset{\mathbf{T}}{\operatorname{max}} \sum_{\tau_i \in \Gamma_S} \omega_i \frac{T_i^{des}}{T_i}, \quad  %\\
\text{Subject to:~  (\ref{eq:period_ns_const_2}), (\ref{eq:period_ns_const_rm}),  (\ref{eq:period_const_1})} \nonumber.
\end{align}
\end{subequations}
\end{myoptimizationproblem}
Although it is non-trivial to solve the above non-linear non-convex optimization problem in its current form, it is possible to transform $\mathbf{P}\mathbf{\ref{opt:period_adapt_ns}}$  into a convex optimization problem using an approach similar to that presented in this paper. However, the analysis using the above formulation is limited by RM bound and also the security task's periods need to satisfy the constraint in Eq. (\ref{eq:period_ns_const_rm}) to follow RM priority order. In addition, rather than only focusing on optimizing the periods of the security tasks, we aim to design a \textit{unified} framework that can achieve other security aspects (\eg responsiveness and atomicity). Hence, instead of simply running security tasks by themselves in idle-time, we propose using a \textit{server} to execute the security tasks. With this approach, for instance, if better responsiveness is desired from security mechanisms, we could increase the priority of the server and allow the server to execute until the security task finishes its desired checking\footnote{This issue is discussed further in Section \ref{sec:lim_n_dis}.}. Not only will the server abstraction allow us to provide better isolation between real-time and security tasks; but it also enables us to integrate additional security properties and provide better execution frequency for certain conditions as we discuss in Sections \ref{sec:evaluation} and \ref{sec:lim_n_dis}.

\section{The Security Server} \label{sec:sec_server}

%Since our goal is to provide security without altering the real-time scheduling, (and hence we refer to as \textit{opportunistic monitoring}), we consider a \textit{lowest-priority server}\footnote{Although it is possible to execute the security tasks  in a fixed-priority order \textit{without} any server, the server abstraction allows us to provide better isolation between real-time and security tasks; and as we have discussed in Section \ref{sec:lim_n_dis}, this framework leads us to incorporate additional security requirements those mentioned in the Introduction.} %Section \ref{subsec:sec_task}.} 
%will be used to execute the security tasks. 

The security server is an abstraction that provides execution time
to the security tasks, according to a preemptive fixed-priority scheduling algorithm (\eg RM). 
%With the inclusion of the server, the overall system is therefore scheduled by a hierarchical scheduling policy; where in the top-level the real-time tasks and the server will be scheduled by a preemptive fixed-priority scheduler, and in the bottom-level, the server will schedule the security tasks according to preemptive fixed-priority order. 
%The concept of server has originally been developed for minimizing the response time of aperiodic tasks that are scheduled together with hard real-time tasks \cite{server_ex1}, and also for providing resource reservation mechanisms \cite{server_ex2,  periodic_server_1}. 
%Many server algorithms (\eg Polling Server \cite{polling_server}, Deferrable Server \cite{deferrable_server}, Sporadic Server \cite{sporadic_server} etc.) have been proposed in the literature for fixed-priority systems. Since many of these mechanisms provide similar guarantees, instead of focusing any particular server algorithm, we consider a generic server abstraction model.  
Instead of focusing on any particular server algorithm, we consider a generic server abstraction model. The security server $\mathcal{S}(Q, P)$ is characterized by the \textit{capacity} $Q$ and \textit{replenishment period} $P$ and works as follows.

%\begin{enumerate}[i.]
%\item ii
%\end{enumerate}

The server may be in two states, \eg \textit{active} and \textit{inactive} and executed with \textit{lowest-priority}. If any security task is activated at time $t$ and if the server is inactive, then the server will become active with capacity $Q$ and relative deadline (\eg next replenishment time) is set as $t + P$. If the server is already active, then the current capacity and relative deadline remain unchanged. When the server is being scheduled, it executes the security tasks according to its own scheduling policy; which we consider fixed-priority RM scheduling in this work. While a security task is executing, the current available capacity is decremented accordingly. The server can be preempted by the scheduler in order to serve the real-time tasks. When the server is preempted, the currently available capacity is no longer decremented. If the available capacity becomes zero and some security task has not yet finished, then the server is suspended until its next replenishment time. Let $t'$ be that replenishment time. At time $t'$, the server is recharged to its full capacity $Q$, the next replenishment time is set as $t' + P$ and the server can execute again. When the last security task has finished executing
and there is no other pending task in the server, the
server will be suspended. Also, the server will become inactive if there are no security tasks ready to execute. 

There is no strict assumption on the smallest time unit of server parameters, \eg $Q, P \in \mathbb{R}^+$ and the security task releases are not bound to the server \cite{server_ab_uk}. Besides, we assume that there is no no task or server release jitters.

%synchronization or precedence constraints among real-time or security tasks 

%\todo{isn't this a feature of having the server? -- (yes, you're right. commented out that statement --M} 
%and there is no task or server release jitters.

\subsection{Reformulation of Period Adaptation Problem using Server}

Since we execute the security tasks within the server, the constraint in Eq. (\ref{eq:period_ns_const_2}) needs to be revised accordingly to consider the server's available capacity and replenishment period. Let us denote $UB_{\mathcal{S}(Q, P), \Gamma_S}$ as the utilization bound for the set of security task $\Gamma_S$ executing within the server with capacity $Q$ and replenishment period\footnote{The calculation of  capacity and replenishment period is discussed in Section \ref{sec:server_param_selec}.} $P$. Since the server schedules the security tasks on a fixed-priority order, we can determine the utilization bound $UB_{\mathcal{S}(Q, P), \Gamma_S}$ using the concept similar to that discussed in literature \cite{serverBound_raj}.
 The authors in work \cite{serverBound_raj} represented the utilization bound as a function of number of tasks running under the server; which is essentially derived from the Liu and Layland's utilization bound \cite{Liu_n_Layland1973}.  In particular,  %when all the periods of the security tasks are $T_i \geq 2P-Q$, and
 when the smallest period of the security task is greater than or equal to $3P - 2Q$, the upper bound of the utilization factor for the security tasks is given by
$UB_{\mathcal{S}(Q, P), \Gamma_S} = n \left[ \left( \tfrac{3 - \tfrac{Q}{P}}{3 - 2 \tfrac{Q}{P}} \right)^{\frac{1}{n}} - 1 \right]$ where $n$ is number of security tasks in the set $\Gamma_S$.
 %utilization bound of the security tasks can be represented by the following lemma. 
%\begin{lemma}
%The upper bound of the utilization factor for the set security tasks $\Gamma_S$ with $T_i \geq 3P - 2Q$  running under the server  $\mathcal{S}(Q, P)$ is
%%\begin{equation}
%$UB_{\mathcal{S}(Q, P), \Gamma_S} = n \left[ \left( \tfrac{3 - \tfrac{Q}{P}}{3 - 2 \tfrac{Q}{P}} \right)^{\frac{1}{n}} - 1 \right]$
%%\end{equation}
%where $n$ is number of security tasks in the set $\Gamma_S$.% and the security tasks have sufficiently large period requirement, \eg $T^{des} = \underset{\tau_i \in \Gamma_S}{\operatorname{min}} T_i^{des} \geq 3P - 2Q$.
%\end{lemma}
%\begin{IEEEproof}
%See \cite[Theorem 7]{serverBound_raj}.
%\end{IEEEproof}
Hence we can define the following constraints on utilization bound %constraint
%\begin{equation} \label{eq:period_const_2}
\begin{align} %\label{eq:period_const_2}
\sum_{\tau_i \in \Gamma_S}\frac{C_i}{T_i} &\leq UB_{\mathcal{S}(Q, P), \Gamma_S} =  n \left[ \left( \tfrac{3 - \tfrac{Q}{P}}{3 - 2 \tfrac{Q}{P}} \right)^{\frac{1}{n}} - 1 \right] \label{eq:period_const_2} \\ 
T_i &\geq 3P - 2Q \quad \forall \tau_i \in \Gamma_S. \label{eq:period_const_4}
\end{align}

Therefore, the period optimization problem with presence of server can be presented similar to that of $\mathbf{P}\mathbf{\ref{opt:period_adapt_ns}}$ except the constraints in Eqs. (\ref{eq:period_ns_const_2}) and (\ref{eq:period_ns_const_rm}) will be replaced by the constraints in Eqs. (\ref{eq:period_const_2}) and (\ref{eq:period_const_4}).
%
%\begin{myoptimizationproblem} \label{opt:period_adapt}
%\vspace*{-2.0em}
%\begin{subequations}
%\begin{align}
%\underset{\mathbf{T}}{\operatorname{max}} \sum_{\tau_i \in \Gamma_S} \omega_i \frac{T_i^{des}}{T_i}  \\
%\text{Subject to:~ (\ref{eq:period_const_2}), (\ref{eq:period_const_4}), and (\ref{eq:period_const_1})} \nonumber.
%\end{align}
%\end{subequations}
%\end{myoptimizationproblem}

%\begin{myoptimizationproblem} \label{opt:period_adapt}
%\vspace*{-2.0em}
%\begin{subequations}
%\begin{align}
%\underset{\mathbf{T}}{\operatorname{max}} & \sum_{\tau_i \in \Gamma_S} \omega_i \frac{T_i^{des}}{T_i}  \\
%\text{Subject to:~} \nonumber \\
% \sum_{\tau_i \in \Gamma_S} \frac{C_i}{T_i} &\leq  n \left[ \left( \tfrac{3 - \tfrac{Q}{P}}{3 - 2 \tfrac{Q}{P}} \right)^{\frac{1}{n}} - 1 \right] \\
%T_i^{des} &\leq T_i \leq T_i^{max}  \quad \forall \tau_i \in \Gamma_S
%\end{align}
%\end{subequations}
%\end{myoptimizationproblem}

\subsection{Geometric Programming (GP) Formulation}

Solving the above constraint non-linear problem %$\mathbf{P}\mathbf{\ref{opt:period_adapt}}$ 
is not straightforward and it always involves compromise  of accepting a local instead of global solution. Therefore we reformulate the optimization problem as a GP. The GP reformulation allows us to solve the problem in an efficient way. When the constraints are mutually consistent, the GP solution approach can always finds a globally optimal solution \cite{GP_tutorial}. In what follows, we first introduce the GP framework and then show how the period adaptation problem can be cast as a GP.

\subsubsection{Geometric Programming}

A non-linear optimization problem can be solved by GP if the problem is formulated as follows \cite{GP_tutorial}
\begin{subequations}
\begin{align}
\underset{\mathbf{X}}{\operatorname{min}}~ & f_0(\mathbf{x})  \\
\text{Subject to:~} %\nonumber \\
 f_i(\mathbf{x}) &\leq  1 \quad i = 1, \cdots, z_p \\
 g_i(\mathbf{x}) &=  1 \quad i = 1, \cdots, z_m
\end{align}
\end{subequations}
where $\mathbf{x} = [x_1, x_2, \cdots, x_z]^{\mathsf{T}}$ denotes the vector of $z$ optimization variables. The functions $f_0(\mathbf{x}), f_1(\mathbf{x}), \cdots, f_{z_p}(\mathbf{x})$ are \textit{posynomial} and $g_1(\mathbf{x}), \cdots, g_{z_m}(\mathbf{x})$ are \textit{monomial} functions, respectively. A function  $g_i(\mathbf{x})$ is  monomial if it can be expressed as
%\begin{equation}
$g_i(\mathbf{x}) = c_i \prod\limits_{l = 1}^{L_i} x_l^{a_l}$
%\end{equation}
where $c_i \in \mathbb{R}^+$ and $a_l \in \mathbb{R}$. Note that the coefficient $c_i$ must be non-negative but the exponents $a_l$ can be any real number including fractional and negative. A posynomial function is the sum of the monomials, and thus can be represented as
%\begin{equation}
$f_i(\mathbf{x}) = \sum\limits_{l=1}^{L_i} c_l x_1^{a_{1l}} x_2^{a_{2l}} \cdots x_z^{a_{1l}}$
%\end{equation}
where $c_l \in \mathbb{R}^+$ and $a_{jl} \in \mathbb{R}$. The posynomials are closed under addition, multiplication and non-negative scaling where the monomials are closed under multiplication and division. %In summary, in order to form a GP problem, the objective function and inequality constraints must be in posynomial forms, and the equality constraints can only be in  monomial forms. 

\subsubsection{Period Adaptation as a GP} \label{subsec:period_gp}

In the following we reformulate the period adaptation problem  %$\mathbf{P}\mathbf{\ref{opt:period_adapt}}$ 
as a GP.

%\begin{lemma}
%\begin{proposition} \label{prop:posy1}
\begin{observation} \label{prop:posy1}
The fundamental measures, \eg $\sum\limits_{\tau_i \in \Gamma_S} \omega_i \frac{T_i^{des}}{T_i} = \sum\limits_{\tau_i \in \Gamma_S} \omega_i T_i^{des} T_i^{-1}$  and $\sum\limits_{\tau_i \in \Gamma_S} \frac{C_i}{T_i} = \sum\limits_{\tau_i \in \Gamma_S} C_i T_i^{-1}$ in the period adaptation problem % $\mathbf{P}\mathbf{\ref{opt:period_adapt}}$ 
are posynomials.
\end{observation}
%\end{proposition}
%\end{lemma}
%\begin{IEEEproof}
This is directly follows from the observation that all the coefficients are non-negative and the variables (\eg periods) are always positive. Besides, we are only summing up positive terms and therefore the terms are closed under addition. Since the requirement for posynomials is that it need to be closed under addition, the above terms are posynomials.

% and the posynomials are closed under addition.
%\end{IEEEproof}

An interesting property of posynomials and monomials is that, if $f(\cdot)$ is a posynomial and $g(\cdot)$ is a monomial, the ratio $\frac{f(\cdot)}{g(\cdot)}$ will become a posynomial. Since $\frac{f(\cdot)}{g(\cdot)}$ is a posynomial, this allows us to express the constraint $f(\cdot) < g(\cdot)$ as $\frac{f(\cdot)}{g(\cdot)} \leq 1$. For example,  we can easily handle the constraint of the form $f(\cdot) \leq \alpha$ where $f(\cdot)$ is a posynomial and $\alpha > 0$. We can refer $\hat{f}(\cdot)$ is an \textit{inverse posynomial} if $\frac{1}{\hat{f}(\cdot)}$ is a posynomial. %In addition, for a posynomoal $f(\cdot)$  and an inverse posynomial  $\hat{f}(\cdot)$ we can expressed constraint of the form 
Besides, we can maximize a non-zero posynomial objective function by minimizing its inverse. Based on the above description, we can reformulate the maximization problem % $\mathbf{P}\mathbf{\ref{opt:period_adapt}}$ 
as a standard GP minimization problem as follows

\begin{myoptimizationproblem} \label{opt:period_adapt_gp}
\vspace*{-2.0em}
\begin{subequations}
\begin{align}
\underset{\mathbf{T}}{\operatorname{min}} \sum_{\tau_i \in \Gamma_S} {\omega_i}^{-1}  {(T_i^{des})}^{-1} T_i \hspace*{-3em} \\
%\text{s. t:~} \hspace*{5em} \nonumber \\
\hspace*{-1em} \text{s. t.:~} \big( \sum_{\tau_i \in \Gamma_S} C_i {T_i}^{-1} \big) \cdot \left( UB_{\mathcal{S}(Q, P), \Gamma_S} \right)^{-1}  &\leq 1\\
(3P - 2Q) {T_i}^{-1} &\leq 1  ~\forall \tau_i \in \Gamma_S \\
T_i^{des} {T_i}^{-1} &\leq 1  ~\forall \tau_i \in \Gamma_S \\
{(T_i^{max})}^{-1} T_i &\leq 1  ~ \forall \tau_i \in \Gamma_S
\end{align}
\end{subequations}
\end{myoptimizationproblem}
where $ \left( UB_{\mathcal{S}(Q, P), \Gamma_S} \right)^{-1} = \tfrac{1}{n \left[ \left( \tfrac{3 - \tfrac{Q}{P}}{3 - 2 \tfrac{Q}{P}} \right)^{\frac{1}{n}} - 1 \right]}  $.

The GP formulation $\mathbf{P}\mathbf{\ref{opt:period_adapt_gp}}$ is not a convex optimization problem since the posynomials are not convex functions \cite{GP_tutorial}. However, $\mathbf{P}\mathbf{\ref{opt:period_adapt_gp}}$ can be converted into a convex optimization problem using logarithmic transformations. The conversion $\mathbf{P}\mathbf{\ref{opt:period_adapt_gp}}$ into a convex optimization problem is based on a logarithmic change of variables, as well as a logarithmic transformation of objective and constraint functions. Instead of using optimization variable $T_i$ let us use the logarithms, \eg $\tilde{T}_i = \log T_i$ and hence $T_i = e^{\tilde{T}_i}$. Besides, let us replace inequality constraints of the form $f_i(\cdot) \leq 1$ with $\log f_i(\cdot) \leq 0$. Using this transformation we can express $\mathbf{P}\mathbf{\ref{opt:period_adapt_gp}}$ as follows

\begin{myoptimizationproblem} \label{opt:period_adapt_gp_convex}
\vspace*{-2.0em}
\begin{subequations}
\begin{align}
\underset{\mathbf{\tilde{T}}}{\operatorname{min}} ~ \log  \sum_{\tau_i \in \Gamma_S} e^{{\omega_i}^{-1}   {(T_i^{des})}^{-1} \tilde{T}_i} \hspace*{-2em} \\
%\text{Subject to:~} \hspace*{5em} \nonumber \\
\hspace*{-1em} \text{s. t.:~} \log e^{  \big( \sum\limits_{\tau_i \in \Gamma_S} C_i {\tilde{T}_i}^{-1} \big) \cdot \left( UB_{\mathcal{S}(Q, P), \Gamma_S} \right)^{-1}  }  &\leq 0\\ \label{eq:ub_bound_gp_convex}
\log e^{ (3P-2Q) {\tilde{T}_i}^{-1} } &\leq 0  ~ \forall \tau_i \in \Gamma_S \\
\log e^{ T_i^{des} {\tilde{T}_i}^{-1} } &\leq 0  ~\forall \tau_i \in \Gamma_S \\
\log e^{{(T_i^{max})}^{-1} \tilde{T}_i} &\leq 0  ~ \forall \tau_i \in \Gamma_S
\end{align}
\end{subequations}
\end{myoptimizationproblem}
where $\tilde{T}_i = \log T_i$ and $\mathbf{\tilde{T}} = [ \tilde{T}_1, \tilde{T}_2, \cdots \tilde{T}_n]^{\mathsf{T}}$. This logarithomic transformation leads $\mathbf{P}\mathbf{\ref{opt:period_adapt_gp_convex}}$ to a convex optimization problem with respect to new variable $\mathbf{\tilde{T}}$ \cite{GP_tutorial} \cite[Secs. 4.5 and 3.1.5]{boyd_book}. Since $\mathbf{P}\mathbf{\ref{opt:period_adapt_gp_convex}}$ is a convex optimization problem, it can be solved using standard algorithms such as \textit{interior-point} method that is known to be solvable in polynomial time \cite[Ch. 11]{boyd_book}. %with a complexity of $O(|\mathbf{\tilde{T}}|^3)$. 

%The following theorem illustrates how $\mathbf{P}\mathbf{\ref{opt:period_adapt_gp}}$ can be converted into a convex optimization problem using logarithmic transformation.

%\begin{theorem}
%$\mathbf{P}\mathbf{\ref{opt:period_adapt_gp}}$ can be transformed into convex optimization problem. 
%\end{theorem}

\section{Selection of the Server Parameters} \label{sec:server_param_selec}

The period adaptation problem in the preceding section is derived based on a given server parameter $\mathcal{S}(Q, P)$ \eg the utilization bound $UB_{\mathcal{S}(Q, P), \Gamma_S}$ in  constraint (\ref{eq:ub_bound_gp_convex}). However, one fundamental problem is to find a suitable pair of server capacity $Q$ and replenishment period $P$ that respects the real-time constraints of the tasks in the system. In this section, for a given period of the security tasks, we formulate a GP problem to determine the server parameters. 

\subsection{Linear Lower-Bound Supply Function}

%
%\begin{figure}[!t]
%\centering
%\includegraphics[width=3.5in]{sbf}
%\caption{Supply Bound Function.}
%\label{fig:sbf}
%\end{figure}

For the security server  with unknown capacity $Q$ and replenishment period $P$, we can derive the lower (upper) bound of $Q$ ($P$) that makes security tasks $\tau_i \in \Gamma_S$ running under server schedulable by using periodic server model introduced in literature \cite{periodic_server_1} \cite{periodic_server_qp} \cite{mn_gp}. The key idea from previous work is that a task $\tau_i$ can be schedulable if minimum supply for the server can match the maximum workload generated by $\tau_i$ and $hp(\tau_i)$ during a time interval $t$. If the server task $\tau_{\mathcal{S}}$ is scheduled by a fixed-priority scheme, the minimum supply of  the server is delivered to the security tasks when its ($k-1$)-th execution has just finished with minimum interference from the high-priority real-time tasks $\tau_j \in \Gamma_R$. Then, the subsequent executions of $k$-th release are maximally delayed by the higher-priority real-time tasks. % as illustrated in Fig \ref{fig:sbf}. 
For this minimum supply, we can parameterize the \textit{linear lower-bound supply function} $\mathbf{lsbf}_{\mathcal{S}}(t)$ with the period and WCET of higher-priority real-time tasks. %as follows. 

The worst-case response time of the server is the longest time from the server being replenished to its capacity being exhausted with the maximum interference from the high-priority real-time tasks, given that there are security tasks ready to use all of the server's available capacity. In order to calculate exact response time of the server, we can use the formula introduced in the work \cite{res_time_1}. Using this exact method, we can calculate the maximum possible preemption on the server from the higher-priority real-time tasks for a certain length of window and add up the server's capacity. The calculation is repeated iteratively by increasing the window size until the window size exceeds the server's relative replenishment period (in this case the system is determined to be unschedulable) or until the window size is stable. Then, the window size is the \textit{busy period} and let us denote it as $w_\mathcal{S}$. 

The worst case release pattern of server occurs when $\tau_{\mathcal{S}}$ and $hp(\tau_{\mathcal{S}})$ is released simultaneously.  The worst-case busy period $w_\mathcal{S}$ is the maximum time duration that the server can take to execute full capacity $Q$ when it is released simultaneously with the higher-priority real-time tasks, $hp(\tau_{\mathcal{S}})$ at the $k$-th release. Therefore, by using the traditional exact analysis \cite{res_time_1} the worst-case busy period can be obtained as 
\begin{equation} \label{eq:busy_period}
w_{\mathcal{S}}^{k+1} = Q + \sum_{\tau_h \in hp(\tau_{\mathcal{S}})} \left\lceil \frac{w_{\mathcal{S}}^{k}}{T_h} \right\rceil \cdot C_h
\end{equation}
where $w_{\mathcal{S}}^{0} = Q$ and $w_{\mathcal{S}} = w_{\mathcal{S}}^{k+1} = w_{\mathcal{S}}^{k}$ when it converges for some $k$. Therefore the worst-case delay at the $k$-th release and thereafter can be represented as 
\begin{equation} \label{eq:del_s_exact}
\Delta_{\mathcal{S}} = \sum_{\tau_h \in hp(\tau_{\mathcal{S}})} \left\lceil \frac{w_{\mathcal{S}}}{T_h} \right\rceil \cdot C_h.
\end{equation}

However, such iterative methods are only amenable to brute-force approach. This is because, the ceiling function with unknown value (\eg the busy period) can not be in our formulation. Thus we take a different approach by approximating $\Delta_{\mathcal{S}}$. During a time interval of $P$, the maximum workload generated by the server and higher-priority real-time tasks can be represented by
\begin{equation} \label{eq:del_s_apprx}
w_{\mathcal{S}} = Q + \sum_{\tau_h \in hp(\tau_{\mathcal{S}})} \left\lceil \frac{P}{T_h} \right\rceil \cdot C_h.
\end{equation}
Thus using Eq. (\ref{eq:del_s_apprx}), we can avoid the iterative calculation by assuming the number of invocation of higher-priority real-time tasks during $P$, not during the exact busy period of the server. Since $\lceil y \rceil \leq y+1$, we linearize $w_{\mathcal{S}}$ by removing the ceiling function and represent Eq.  (\ref{eq:del_s_apprx}) as 
%\begin{equation}
$w_{\mathcal{S}} = Q + \sum\limits_{\tau_h \in hp(\tau_{\mathcal{S}})} \left( \frac{P}{T_h} + 1 \right) \cdot C_h$.
%\end{equation}
Therefore, the worst-case linear lower-bound supply function of the security server during a time interval $t$ \cite{periodic_server_1} is given by 
\begin{equation} \label{eq:lsbf_final}
\mathbf{lsbf}_{\mathcal{S}}(t) = \frac{Q}{P} \left[ t - (P - Q) - \Delta_{\mathcal{S}}  \right]
\end{equation}
where $\Delta_{\mathcal{S}} = \sum\limits_{\tau_h \in hp(\tau_{\mathcal{S}})} \left( \frac{P}{T_h} + 1 \right) \cdot C_h$.

\subsection{Sufficient Bound for Schedulability of the Security Tasks} \label{subsec:sec_task_bound}

In order to derive the minimum capacity that guarantees to schedule $\tau_i \in \Gamma_S$, let us consider the situation when $\tau_i$ barely meets it deadline at $t = D_i$ with the worst-case interference from high-priority security tasks, $hp(\tau_i) \in \Gamma_S$. Let us now define the \textit{critical instant}%\footnote{Although we do not consider task jitters, the presented analysis can  similarly be applied to cases with jitters without loss of generality. As an example, the worst-case situation of $\tau_i$ is when $hp(\tau_i)$ have experienced their maximum jitters and are released at the same time with $\tau_i$.} 
 of the security tasks, \eg the worst-case response time of $\tau_i$ when $\tau_i$ and $hp(\tau_i)$ are released simultaneously at the end of server's ($k-1$)-th execution and suffer worst-case preemptions from $k$-th release and thereafter \cite{mn_gp}. Let us denote $I_i$ as the worst-case workload generated by the $\tau_i$ and $hp(\tau_i)$ from critical instant to deadline of $\tau_i$ given by
\begin{equation}
I_i = C_i + \sum\limits_{\tau_h \in hp(\tau_i)} \left\lceil \frac{D_i}{T_h} \right\rceil \cdot C_h.
\end{equation} 
In order to ensure the schedulability of the security task $\tau_i$, the minimum supply delivered by the server has to be greater than or equal to the worst-case workload during the time interval $D_i$, \eg  %Mathematically which can be expressed as,
\begin{equation} \label{eq:sh_bound_sec}
\mathbf{lsbf}_{\mathcal{S}}(D_i) \geq I_i \quad \forall \tau_i \in \Gamma_S
\end{equation}
where $\mathbf{lsbf}_{\mathcal{S}}(\cdot)$ is given by Eq. (\ref{eq:lsbf_final}).

It is worth noting that Eq. (\ref{eq:sh_bound_sec}) is only a sufficient and not necessary condition. The security task $\tau_i$ can be schedulable if and only if there exists a time instance $t \leq D_i$ such that the inequality in Eq. (\ref{eq:sh_bound_sec}) holds. However, we use the sufficient condition in Eq. (\ref{eq:sh_bound_sec}) because the presence of time in the necessary condition makes the proposed optimization framework inapplicable to the problem under consideration. Despite the fact that this bound may not be exact and may incur approximation error in the supply function, as we show in Sections \ref{sec:evaluation} and \ref{sec:lim_n_dis}, it enables us to ensure opportunistic execution of the security tasks without violating real-time constraints.

\subsection{Formulation as an Optimization Problem}

In the following we present the optimization problem formulation to determine the server parameters $Q$ and $P$.

\subsubsection{Objective Function} 
The objective of periodic monitoring in the system is to ensure maximal processor utilization for the security tasks, without violating the real-time constraints of the system. Therefore, we define the objective function as follows
\begin{equation} \label{eq:ser_obj}
\underset{Q, P}{\operatorname{max}}~ \frac{Q}{P} 
\end{equation}
where the server parameters, $Q$ and $P$ are the optimization variables.

\subsubsection{Server Schedulability Constraint}

 A server is schedulable if worst-case response time of the server does not exceed its replenishment period \cite{server_ab_uk}. Hence we represent the server schdulability constraint as follows
 \begin{equation} \label{eq:ser_con1}
 Q + \sum\limits_{\tau_h \in hp(\tau_{\mathcal{S}})} \left( \frac{P}{T_h} + 1 \right) \cdot C_h \leq P.
 \end{equation}

\subsubsection{Server Bound Constraints}

As we have discussed in Section \ref{subsec:sec_task_bound}, in order to guarantee schedulability of each of the security tasks  $\tau_i \in \Gamma_S$, the linear lower-bound supply function must be greater than or equal to  worst-case workload during the time interval $D_i$. Therefore, from Eq. (\ref{eq:sh_bound_sec}) the constraints on the server supply bound to ensure schedulability of the security tasks  can be expressed as
\begin{equation} \label{eq:ser_con2}
 \frac{Q}{P} \left[ D_i - (P - Q) - \Delta_{\mathcal{S}}  \right] \geq I_i \quad \forall \tau_i \in \Gamma_S 
\end{equation}
where $T_i^{des} \leq D_i = T_i^* \leq T_i^{max}$ and $T_i^*$ is obtained by solving the period adaptation problem $\mathbf{P}\mathbf{\ref{opt:period_adapt_gp_convex}}$. Notice that, in Eq. (\ref{eq:ser_con2}) $I_i = C_i + \sum\limits_{\tau_h \in hp(\tau_i)} \left\lceil \frac{T_i^*}{T_h} \right\rceil \cdot C_h$ is the worst-case workload generated by $\tau_i$ and $hp(\tau_i)$ during the time interval of $T_i^*$. This is a constant for a given input.

%Using the objective function in (\ref{eq:ser_obj}), and the constraints in (\ref{eq:ser_con1}) and (\ref{eq:ser_con2}), the server parameter selection problem can be expressed as follows.

%\begin{myoptimizationproblem} \label{opt:server_non-lin}
%\vspace*{-2.0em}
%\begin{subequations}
%\begin{align}
%\underset{Q, P}{\operatorname{max}}~ \frac{Q}{P}  \hspace*{1em} \label{eq:obj_server_nonlin}\\
%\text{Subject to:~} \hspace*{5em} \nonumber \\
% Q + \sum\limits_{\tau_h \in hp(\tau_{\mathcal{S}})} \left( \frac{P}{T_h} + 1 \right) \cdot C_h &\leq P \label{eq:con1_server_nonlin}\\ 
% \frac{Q}{P} \left[ T_i^* - (P - Q) - \Delta_{\mathcal{S}}  \right] &\geq I_i \quad \forall \tau_i \in \Gamma_S  \label{eq:con2_server_nonlin}
%\end{align}
%\end{subequations}
%\end{myoptimizationproblem}

%\begin{myoptimizationproblem} \label{opt:server_non-lin}
%\vspace*{-2.0em}
%\begin{subequations}
%\begin{align}
%\underset{Q, P}{\operatorname{max}}~ \frac{Q}{P}  \hspace*{1em} \label{eq:obj_server_nonlin}\\
%\text{Subject to:~   (\ref{eq:ser_con1}), (\ref{eq:ser_con2})}.
%\end{align}
%\end{subequations}
%\end{myoptimizationproblem}

\subsection{Geometric Programming Formulation}

%Since  $\mathbf{P}\mathbf{\ref{opt:server_non-lin}}$ is non-linear constraint optimization problem, we reformulate it as a GP problem.

\begin{proposition} \label{prop:posy2}
The objective function in %(\ref{eq:obj_server_nonlin}) 
Eq. (\ref{eq:ser_obj}) and the constraint in %(\ref{eq:con1_server_nonlin}) 
Eq. (\ref{eq:ser_con1}) can be expressed in posynomial form.
\end{proposition}
%\end{lemma}
\begin{IEEEproof}
The proof follows by rearranging the terms in posynomial form and transform the objective function in Eq. (\ref{eq:ser_obj}) into minimization expression. For details refer to Appendix \ref{appsec:proof_prop}.
\end{IEEEproof}

With a view to expressing the server bound constraint as posynomial form, we can rearrange %(\ref{eq:con2_server_nonlin})
Eq. (\ref{eq:ser_con2}) as follows
\begin{equation} \label{eq:server_bound_npoys}
\frac{P (Q + I_i) + \Delta_{S} Q}{Q (Q + T_i^*)} \leq 1 \quad \forall \tau_i \in \Gamma_S.
\end{equation}
Recall that, in order to represent the constraint of the form $\frac{f(\cdot)}{g(\cdot)} \leq 1$ the denominator must be monomial. However, the inequality in (\ref{eq:server_bound_npoys}) does not conform to a posynomial form due to the posymolial term in the denominator, \eg $Q (Q + T_i^*) = Q^2 + Q T_i^*$. The following theorem illustrates how the constraints on server bound can be represented in posymonial form.

%Hence we use the geometric mean approximation \cite[Ch. 2]{gp_comm} and represent the server bound constraint in posymonial form as given by the following thereom.

\begin{theorem} \label{th:posy}
The server bound constraints can be formulated as the following posynomial form
\begin{equation}
\left[P  (Q + I_i) + \Delta_{S}  Q \right] \cdot {\left[Q \cdot \hat{g}(Q, T_i^*) \right]}^{-1} \leq 1 \quad \forall \tau_i \in \Gamma_S.
\end{equation}
\end{theorem}
\begin{IEEEproof}
The theorem is proved by using the geometric mean approximation \cite[Ch. 2]{gp_comm} of posynomials. For detailed proof refer to Appendix \ref{appsec:proof_thm}.
\end{IEEEproof}

Using the logarithmic transformation presented in Section \ref{subsec:period_gp}, we  can formulate server parameter selection problem as a GP in convex form as follows

\begin{myoptimizationproblem} \label{opt:server_gp_convex}
\vspace*{-2.0em}
\begin{subequations}
\begin{align}
\underset{\tilde{Q}, \tilde{P}}{\operatorname{min}}~ \log e^{\tilde{Q}^{-1} \tilde{P}}  \hspace*{1em} \\
%\text{Subject to:~} \hspace*{5em} \nonumber \\
\hspace*{-1.3em} \text{s. t.:~} \log e^{\big[\tilde{Q} + \sum\limits_{\tau_h \in hp(\tau_S)} (\tilde{P} + T_h) \cdot T_h^{-1} \cdot C_h \big] \cdot \tilde{P}^{-1} }  &\leq 0 \\
 \log e^{\left[\tilde{P}  (\tilde{Q} + I_i) + \Delta_{S}  \tilde{Q} \right] \cdot {\left[\tilde{Q} \cdot \hat{g}(\tilde{Q}, T_i^*) \right]}^{-1}} &\leq 0 ~ \forall \tau_i \in \Gamma_S \label{eq:gp_server_convex_mono_approx}
\end{align}
\end{subequations}
\end{myoptimizationproblem}
where $\tilde{Q} = \log Q$ and $\tilde{P} = \log P$. Since $\mathbf{P}\mathbf{\ref{opt:server_gp_convex}}$ is a convex optimization problem, it is solvable using standard algorithm such as interior-point method. 

%the server parameter selection problem as the following convex optimization problem

%
%
%\begin{myoptimizationproblem} \label{opt:server_opt}
%\vspace*{-2.0em}
%\begin{subequations}
%\begin{align}
%\operatorname{min} f_0(Q, P)  =  Q^{-1} P \\
%\text{Subject to:~} \nonumber \\
%(Q + \Delta_{S}) P^{-1} \leq 1 \\
%{\left(P \cdot (Q + I_j) + \Delta_{S} \cdot Q) \cdot (Q \cdot \hat{g}(Q, d_j)\right)}^{-1} \leq 1 \quad \forall j
%\end{align}
%\end{subequations}
%\end{myoptimizationproblem}
%where 
%\begin{equation}
%\Delta_{S} =
%  \sum_{\tau_h \in hp(\tau_S)} \left( (P + T_h) \cdot T_h^{-1} \cdot C_h \right) 
%\end{equation}
%and
%\begin{equation}
%I_j = C_j + \sum_{\tau_h \in hp(\tau_j)} \left\lceil \frac{D_j}{T_h} \right\rceil C_h
%\end{equation}
%
%

\section{Algorithm Development} \label{sec:algorithm}

We develop an iterative scheme to obtain the period of the security tasks and the server parameters jointly. The overall procedure, as summarized in \textbf{Algorithm \ref{alg:sec_schd}} works as follows. The algorithm starts with the maximum allowable periods (\eg $T_i^{max}$) and try to find the $Q$ and $P$ by solving the server parameter selection problem $\mathbf{P}\mathbf{\ref{opt:server_gp_convex}}$. If there is no solution, which implies that the constraints in  $\mathbf{P}\mathbf{\ref{opt:server_gp_convex}}$ are mutually inconsistent, it is not possible to provide a server for the security tasks without violating the real-time constraints of the system. Since it is then not possible to integrate security tasks in the given system, the algorithm reports the set of real-time and security tasks as unschedulable. %If there is a solution, estimate the best period for the security tasks by solving $\mathbf{P}\mathbf{\ref{opt:period_adapt_gp_convex}}$. If no solution found, which means the constraints are mutually

If there is a solution, we iteratively estimate the best period for the security tasks (Lines 8-26). In particular, for the given server parameter $Q(j)$ and $P(j)$ for any iteration $j$, the algorithm estimates the periods for next iteration by solving $\mathbf{P}\mathbf{\ref{opt:period_adapt_gp_convex}}$. Likewise, for a given period vector, we calculate the best server parameters that make the constraints on $\mathbf{P}\mathbf{\ref{opt:server_gp_convex}}$ mutually consistent. Let $\eta(j)$ be the objective value by solving $\mathbf{P}\mathbf{\ref{opt:period_adapt_gp_convex}}$ at iteration $j$.  The iteration is repeated as long as the difference of objective value in successive iterations is greater than some predefined tolerance for convergence $\epsilon$, \eg $|\eta(j) - \eta(j-1)| > \epsilon$ or the maximum allowable iteration counter $J_{max}$ is not exceeded. From our experiments we find that, the algorithm generally converged within $3\text{-}5$ iterations for most of the schedulable task-sets with tolerance $\epsilon = 10^{-16}$.

Recall that, the approximation quality of $\hat{g}(\tilde{Q}, T_i^*)$ [\eg constraint (\ref{eq:gp_server_convex_mono_approx}) in $\mathbf{P}\mathbf{\ref{opt:server_gp_convex}}$] depends on the choice of $y_0$. Thus, in the optimization procedure (\eg Line 3 and 18), we iteratively approximate $\hat{g}(\tilde{Q}, T_i^*)$ by updating $a$ and $b$ according to the intermediate solution of $\tilde{Q}$. That is, until the objective value converges, we use $\tilde{Q}$ at $k$-th step as $y_0$ at $(k + 1)$-th step. In our experiments, we choose the initial value of $y_0$ as $1$, and the objective value of $\mathbf{P}\mathbf{\ref{opt:server_gp_convex}}$ converged  within $5$ iterations.

\begin{algorithm}[H]

\newcommand{\algorithmicbreak}{\textbf{break}}
\newcommand{\BREAK}{\STATE \algorithmicbreak}
\renewcommand\algorithmiccomment[1]{%
 {\it /* {#1} */} %
}
\renewcommand{\algorithmicrequire}{\textbf{Input:}}
    \renewcommand{\algorithmicensure}{\textbf{Output:}}

\begin{algorithmic}[1]
\begin{footnotesize}
\REQUIRE Set of real-time and security tasks, $\Gamma_R$ and $\Gamma_S$, respectively
    \ENSURE The tuple $\left\lbrace \mathbf{T}^*, Q^*, P^*\right\rbrace$, \eg periods of the security tasks and the server parameters if the task-set is schedulable; $\mathsf{Unscheduable}$	 otherwise
\vspace{0.2em}
%\vspace*{-0.5em}

%\STATE Initialize Security tasks $\tau_i = (C_i, \widehat{T}_i)$ 
%\STATE Solve the top level GP to obtain server parameters
%
% \IF {$\mathsf{Soulution Found}$}
%        \STATE {\tiny\textit{/* Possible to integrate Security tasks with desired period*/}}
%        \STATE \textbf{return} Server parameters $(Q^*, P^*)$
% \ELSE
\STATE Initialize $j : = 1$
 \STATE Initialize security task's period vector $\mathbf{T}(j) := [T_i^{max}]_{\forall \tau_i \in \Gamma_S}^{\mathsf{T}}$ 
 \STATE For the given $\mathbf{T}(j)$, Solve  $\mathbf{P}\mathbf{\ref{opt:server_gp_convex}}$ to obtain server parameters
 \IF {no solution found}
 				\STATE \COMMENT{not possible to integrate security tasks in the system}
				\STATE \textbf{return} $\mathsf{Unscheduable}$	
				\ELSE	 		

% \REPEAT  
%\STATE \COMMENT $Q_*, P_*$ be the solution from $\mathbf{P}\mathbf{\ref{opt:server_gp_convex}}$

\STATE Set $Q(j) = Q_*$, $P(j) = P_*$ where $Q_*, P_*$ be the solution from $\mathbf{P}\mathbf{\ref{opt:server_gp_convex}}$

\WHILE{period perturbations are not minimized \AND $j < J_{max}$}
		
				%\STATE Solve GP to obtain server parameters 
				\STATE Update $j := j+1$
				\STATE Solve the period adaptation problem $\mathbf{P}\mathbf{\ref{opt:period_adapt_gp_convex}}$ using server parameters  $Q(j-1), P(j-1)$
					\IF {no solution found}
					\STATE \COMMENT{return current best solution}
					
				\STATE Set $\mathbf{T}(j):= \mathbf{T}(j-1), Q(j) := Q(j-1), P(j) := P(j-1)$	
				\BREAK
				\ENDIF				
				
				\STATE Update the period vector $\mathbf{T}(j) := \mathbf{T}_{*}$  where $\mathbf{T}_{*}$ is the solution obtained from $\mathbf{P}\mathbf{\ref{opt:period_adapt_gp_convex}}$
				
				%to minimize perturbation
				\STATE For the given $\mathbf{T}(j)$, Solve $\mathbf{P}\mathbf{\ref{opt:server_gp_convex}}$ to obtain server parameters 
				
				\IF {no solution found}
				\STATE \COMMENT{return current best solution}
				\STATE Set $Q(j) := Q(j-1), P(j) := P(j-1)$	
				\BREAK
				\ENDIF

%		 	\IF {$\mathsf{Soulution Not Found}$}
%				\STATE \textbf{return} unscheduable		 		
%		 	\ELSE
		 		%\STATE {\tiny\textit{/*Not possible to integrate Security tasks in the current system*/}}
	 		\STATE Update the server parameters $Q(j) := Q_*, P(j) := P_*$

%		 		\STATE \textbf{return} best found solution \OR unscheduable
		 		%\ENDIF
		 		%\UNTIL{$\mathsf{Soulution Found}$ \AND Security perturbations are not minimized}

		 		\ENDWHILE
		 				 		\STATE \textbf{return} $\left\lbrace \mathbf{T}^* := \mathbf{T}(j), Q^* := Q(j), P^* := P(j) \right\rbrace$
		 				 	\ENDIF

    %\ENDIF

\end{footnotesize}

\end{algorithmic}
\caption{Security Period and Server Parameter Selection}
\label{alg:sec_schd}
\end{algorithm}

\section{Evaluation} \label{sec:evaluation}

To evaluate the performance of our proposed opportunistic monitoring model, we performed experiments using randomly generated workloads. The parameters used in our evaluation are summarized in Table \ref{tab:ex_param}.

 %We had XXX main objectives for our evaluation: 

\begin{table}[!thb]
\renewcommand{\arraystretch}{1.3}
\caption{Experimental Parameters}
\label{tab:ex_param}
\centering
%\begin{tabular}{c||p{5.9cm}}
\begin{tabular}{p{4.9cm}||c}
\hline %\\
\bfseries Parameter & \bfseries Values\\
\hline\hline
Number of real-time tasks, $m$ & $[3, 10]$ \\
Number of security tasks, $n$ & $[2, 5]$ \\
Real-time task period, $T_i$ & $[10ms, 100ms]$ \\
Desired period for security tasks, $T_i^{des}$ & $[250ms, 500ms]$\\
Maximum allowable period, $T_i^{max}$ & $[5000ms, 5050ms]$\\
%Desired period for security tasks, $T_i^{des}$ & $0.5T_i^{max}$ \\
Tolerance for convergence, $\epsilon$ & $10^{-16}$\\
Maximum replenishment period (for exhaustive search), $P_{max}$ & $2500$ \\
Exhaustive search granularity, $\delta$ & $0.5$ \\
\hline
\end{tabular}
\end{table}

\subsection{Evaluation Setup}

The base-utilization of a task-set is defined as the total sum of the task utilizations. The real-time and security task-sets are grouped by base-utilization from $[0.01+0.1 \cdot i, 0.1+0.1 \cdot i]$ where $i \in \mathbb{Z}$ and the number of base-utilization groups are specified in the corresponding experiments. 
%$0 \leq i \leq 4$ for real-time tasks, and $0 \leq i \leq 3$  for the security tasks, respectively. 
This allows us to generate task-sets with an even distribution of tasks. %Each utilization group contains fixed number of tasks. 
Each task-set instance contains $[3, 10]$ real-time and $[2, 5]$ security tasks. Each real-time task $\tau_i \in \Gamma_R$ has a period $T_i \in [10ms, 100ms]$. 
The desired and maximum allowable periods for the security tasks are selected from $[250ms, 500ms]$ and  $[5000ms, 5050ms]$, respectively. 
%The maximum allowable period $T_i^{max}$ for the security task $\tau_i \in \Gamma_S$ is selected from $[30ms, 1000ms]$ and the desired period is considered as $T_i^{des} = 0.5 T_i^{max}$.

The utilization of the real-time and security tasks are generated by the UUniFast \cite{uunifast} algorithm. For a given utilization $U_i$, the execution time of $\tau_i$ is generated by $C_i = U_i T_i$. %Each utilization group contains $100$ task-set. 
%The priorities for real-time and security tasks are assigned according to the RM \cite{Liu_n_Layland1973} algorithm. %The tolerance for convergence, $\epsilon = 10^{-16}$ and maximum iteration limit, $J_{max}$ is set as $200$.  
We use GGPLAB \cite{ggplab} to solve the GPs. All the experiments are performed on Intel Pentium N3530 2.16 GHz processor with 4 GB RAM.

\subsection{Results}

\subsubsection{Comparison with Exact Method}

We compare our GP-based approach with an exhaustive search method\footnote{Similar search method has also been discussed in literature \cite{periodic_server_1, server_ab_uk, mn_gp}.} based on exact analysis. In this exhaustive search, we assign server replenishment period from $1$ to $P_{max}$ with a granularity of $\delta$. For each period, we determine the minimum capacity requirements that makes the tasks schedulable. From the set of feasible period and capacity pair, we take the pair that maximizes the server utilization. Notice that, the minimum server capacity $Q_{min}(\tau_i, P)$ for $\tau_i \in \Gamma_S$ with a given $P$ can be obtained by solving the quadratic inequality in Eq. (\ref{eq:ser_con2}), which is given by
\begin{equation}
Q_{min}(\tau_i, P) = \tfrac{-(D_i - P - \Delta_{\mathcal{S}}) + \sqrt{{(D_i - P - \Delta_{\mathcal{S}})}^2 + 4 I_i P}}{2}.
\end{equation}
In the above equation $\Delta_{\mathcal{S}}$ is calculated by exact method, \eg $\Delta_{\mathcal{S}} = \sum\limits_{\tau_h \in hp(\tau_{\mathcal{S}})} \left\lceil \frac{w_{\mathcal{S}}}{T_h} \right\rceil \cdot C_h$ where $w_{\mathcal{S}}$ is obtained from Eq. (\ref{eq:busy_period}). In order to find the minimum required capacity of the server for a given replenishment period $P$, we take the maximum of the capacity $Q_{min}(\tau_i, P)$ over all the security tasks $\tau_i \in \Gamma_S$ which is defined as %follows
\begin{equation}
Q_{min}(P) = \underset{\tau_i \in \Gamma_S}{\operatorname{max}} \left\lbrace Q_{min}(\tau_i, P) \right\rbrace.
\end{equation}
Hence any $Q$ from $[Q_{min}(P), P]$ such that $Q + \Delta_{\mathcal{S}} \leq P$ will be the feasible capacity that makes the task-set schedulable.

%\begin{figure}[!t]
%\centering
%\includegraphics[width=3.5in]{ex_vs_gp_solcount_exact3}
%\caption{Normalized percentage of the number of schedulable task-sets. For each utilization group, we randomly generate $100$ task-sets and compare the schedulability of both schemes. For exhaustive search, we set $P_{max} = 2500$ with search granularity $\delta = 0.5$.}
%\label{fig:ex_vs_gp_solcount}
%\end{figure}

\begin{figure}[!t]
\centering
\includegraphics[width=3.0in]{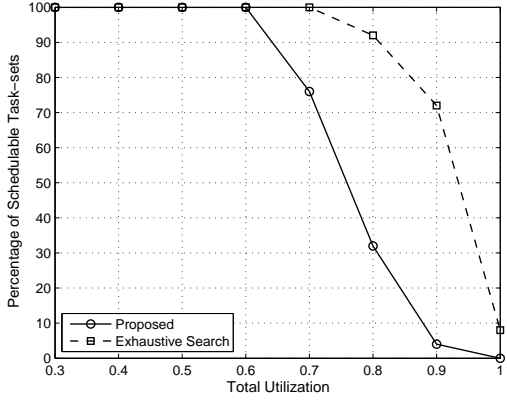}
\caption{Normalized percentage of the number of schedulable task-sets. The base-utilization of the real-time tasks are varied from $[0.01+0.1 \cdot i, 0.1+0.1 \cdot i]$ where $0 \leq i \leq 8, i \in \mathbb{Z}$. The utilizations of the security tasks are generated from $[0.11, 0.20]$. For exhaustive search, we set $P_{max} = 2500$ with search granularity $\delta = 0.5$.}
\label{fig:ex_vs_gp_solcount}
\end{figure}

In Fig. \ref{fig:ex_vs_gp_solcount} we compare the number of schedulable task-sets found in the proposed method and exhaustive search. For exhaustive search, we set $P_{max} = 2500$ with granularity $\delta = 0.5$.  %$x$-axis and $y$-axis of Fig. \ref{fig:ex_vs_gp_solcount} represents the base-utilization of the real-time and security tasks, respectively; and $z$-axis represents normalized percentage of the number of schedulable task-sets, i.e., $\frac{\text{number of schedulable task-sets by exhaustive search}}{\text{number of schedulable task-sets by GP}}$. The closer the ratio to $1$, the closer the number of schedulable task-sets found by GP to that of exhaustive search. 
As we can see from figure, the difference in terms of schchdulable task-sets found by exhaustive search compared to GP increases for higher base-utilization. We can attribute that due to approximation of supply function in the security server. Recall that, the exhaustive search method calculates minimum capacity of the server by exact analysis of the busy period. In contrast, the proposed method  approximates the interference to the server from real-time tasks during the interval of server replenishment period and linearize it by taking the ceiling off. While this approximation error is small for low utilization cases, as the base-utilization increases the error accumulates and reduces schedulability.  However, still it is possible to accumulate task-sets for higher base-utilization.

\begin{figure}[!t]
\centering
\includegraphics[width=3.0in]{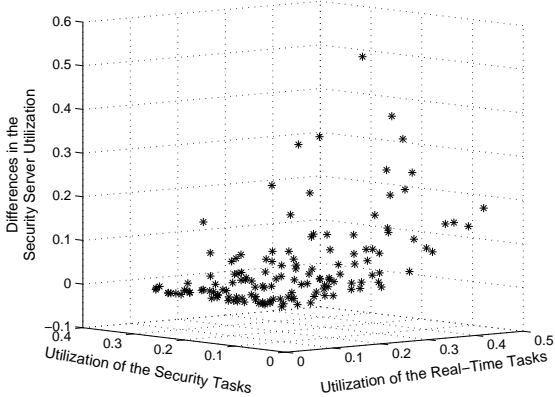}
\caption{Exhaustive search vs. proposed approach: difference in the server utilization for schedulable task-sets. For each utilization group, we randomly generate $100$ task-sets and compare the schedulability of both schemes.}
\label{fig:ex_vs_gp_diff}
\end{figure}

The quality of solution (\eg server utilization) obtained by GP and exhaustive search is illustrated in Fig. \ref{fig:ex_vs_gp_diff}. The $z$-axis in this figure represents the difference in server utilization, \eg $\left(\frac{Q^{\rm EX}}{P^{\rm EX}} - \frac{Q^{\rm GP}}{P^{\rm GP}} \right)$ where $Q^{\rm EX}$ and $Q^{\rm GP}$ ($P^{\rm EX}$ and $P^{\rm GP}$) represent the capacity (replenishment period) obtained from exhaustive search and proposed method, respectively. For low-to-medium utilization cases, the difference is close to zero, which implies the quality of the solution obtained by the GP method is similar to that of obtained by exhaustive search.  However, when the utilization is higher the exhaustive search outperforms the proposed method. Again, we can attribute this due to the approximation of supply function in the security server. 

It is worth noting that the solution obtained by exhaustive search may not be optimal in a sense that the actual replenishment period may appear beyond $P_{max}$. As we can see from Fig. \ref{fig:ex_vs_gp_diff}, for some task-sets the difference is less than zero, \eg $\frac{Q^{\rm EX}}{P^{\rm EX}}$ is lower than $\frac{Q^{\rm GP}}{P^{\rm GP}}$.  We highlight that the actual search region to find the optimal server parameters for exhaustive search may widely vary based on task-set inputs; and  can only be found numerically by trial-and-error. Instead, the proposed method provides a generic approach to analyze the system that is independent of task-set input parameters.

%Notice that, the solving time of each method is not compared due to the fact that 
We note that the proposed GP-based approach can solve a given task-set in \textit{seconds}, while the exhaustive search method generally takes few minutes to couple of hours depending on the size of $P_{max}$ and the search granularity $\delta$. Besides, the exhaustive search method is not scalable for task-sets with large number of tasks.

\subsubsection{Effectiveness of Security}

\begin{figure}[!t]
\centering
\includegraphics[width=3.0in]{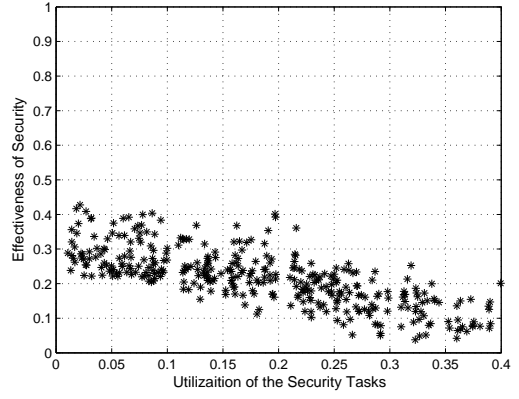}
\caption{The effectiveness of security is measured as %normalized Euclidean distance of period vector, which is given by 
$\xi = \frac{ {\lVert \mathbf{T}^* - \mathbf{T^{des}}\rVert}_2}{{\lVert \mathbf{T^{max}} - \mathbf{T^{des}}\rVert}_2}$. The base-utilization of the real-time tasks are taken from  $[0.31, 0.4]$ and the base-utilization of the security tasks are varied from $[0.01+0.1 \cdot i, 0.1+0.1 \cdot i]$ where $0 \leq i \leq 3, i \in \mathbb{Z}$. Hence the total utilization of the system varied from $[0.3, 0.8]$. Each utilization group contains $100$ task-sets.}
\label{fig:distance}
\end{figure}

In Fig. \ref{fig:distance}, we observe the effectiveness of security of the system by means of tightness of the desired periods, %. Tightness of the each of the task's period is measured by normalized Euclidian distance of the period vector, 
\eg $\xi = \frac{ {\lVert \mathbf{T}^* - \mathbf{T^{des}}\rVert}_2}{{\lVert \mathbf{T^{max}} - \mathbf{T^{des}}\rVert}_2}$ where $\mathbf{T}^*$ is the solution obtained from \textbf{Algorithm \ref{alg:sec_schd}}, $\mathbf{T^{des}} = [T_i^{des}]_{\forall \tau_i \in \Gamma_S}^{\mathsf{T}} $ and $\mathbf{T^{max}} = [T_i^{max}]_{\forall \tau_i \in \Gamma_S}^{\mathsf{T}} $ are the desired and maximum period vector, respectively, and ${\lVert \cdot \rVert}_2$ denotes the Euclidean norm. The closer the value of $\xi$
to $0$, the nearer the period of each of the security task is to the  desired period. Each of the data point in Fig. \ref{fig:distance} represents schedulable task-sets. We find that most cases the algorithm finds the periods that is within $20\%$ of the desired period value.

We note that, our GP approach took $5.33$ second on average with $23.33$ standard deviation to analyze a set of tasks with parameters specified by Table \ref{tab:ex_param}.
%of $2\text{-}15$ tasks. 
Thus, the proposed approach solves problems of reasonable size in \textit{seconds}, which is an acceptable amount of time for \textit{offline analysis}.

\subsubsection{Experiment with a Security Application}

\begin{figure}[!t]
\centering
\includegraphics[width=3.0in]{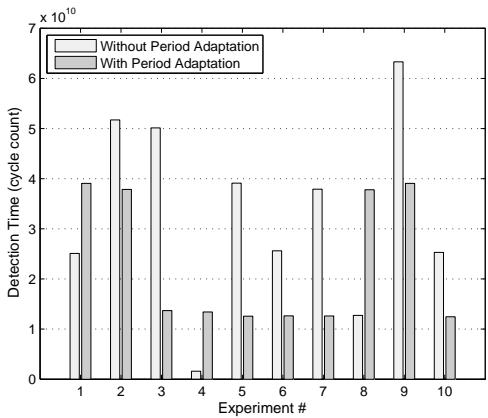}
\caption{Time to detect attacks: when the periods are optimized vs when set to maximum allowable frequency. Each of the bar-groups represents individual experiment run. We use read time-stamp counter (RDTSC) instruction to measure the cycle count.} %Most of the cases adaptive mechanism detect attacks quickly.}
\label{fig:time2detect}
\end{figure}

With a view to observing that whether the security tasks can perform desired checking after period adaptation, we perform experiments with Tripwire in RTAI-patched Linux RTOS. For this, we purposely compromise one of the tasks and launch the attack by modifying the contents in the  $\mathtt{/sbin}$ directory. For each of the experiments run, we start with a clean system state, launch attack at any random point of the task execution and log the cycle count requires to detect the attack by Tripwire. We obtain the periods by solving the period adaptation problem and set it as the period of Tripwire in the RTOS. For non-adaptive case, period of Tripwire is set to $T_i^{max}$. Each of the bars in the x-axis of Fig. \ref{fig:time2detect} denotes individual experiment run, and y-axis shows the corresponding cycle counts to detect the attack. As shown in Fig. \ref{fig:time2detect}, out of 10 experiments we find 3 instances, where the non-adaptive period assignment outperforms. This is mainly because the actual time to detect the attacks depends on when the attack is launched and the corresponding scheduling point of the security routine. However in general case, the proposed period adaptation mechanism provides better monitoring frequency and hence it is more likely to detect breaches as soon the attack is launched.

\section{Limitations and Discussion} \label{sec:lim_n_dis}

While the proposed method provides an approach for integrating security in RTS and a metric to measure the effectiveness of such security integration, there are still some areas for improvement for the mechanisms presented in this paper.  The proposed server-based approach imposes additional constraints on periodic monitoring compared to non-server case and also based on approximation of supply function in the security server which somewhat limits schedulability. However, as we discuss in the following, the proposed frameworks allows us to provide better control in terms of security enforcement.  

Aforesaid, in order to integrate security policy into the system along with  monitoring frequency, other  performance aspects such as responsiveness and atomicity of the security tasks also need to be considered. While the proposed server-based method provides us a first step to ensure periodic monitoring, this approach can be extended to satisfy responsiveness and atomicity properties as well. 

In the case when responsiveness and atomicity of the security mechanism should be ensured, the proposed framework can be modified as follows. Whilst the security task requires better responsiveness and/or needs to perform special atomic operation, the priority of the server can be increased to a priority that is strictly higher than all (or some) of the real-time tasks, depending on the requirements of the security event. Besides, if the security task running under server is not the highest-priority security task, the priority of that task itself is also increased. If the server's capacity is exhausted while executing any atomic operation or fine-grained checking, we allow the server to overrun, \eg the server continues to execute at the same priority until the security checking is completed. When the server overruns, the allocated capacity at the start of the next server replenishment period is reduced by the amount of the overrun.

It is worth mentioning that, the cost of responsiveness and atomicity by means of priority inversion will be reflected by compromising the timing constraints of some of the real-time tasks. In such cases, the schedulability analysis need to be performed considering maximum blocking time of the security events. Besides, The scheduling policy should identify which real-time or security tasks can be dropped to provide better trade-off between control system performance and defense against security vulnerabilities. In addition, depending on the actual implementations of the security routines, the scheduling framework may need to follow certain precedence constraints. Analogous to the task example illustrated in Table \ref{table:rtos}, in order to ensure that the security application has not been compromised, the security application's own binary may be scanned first before checking the system binary or library files. We intend to explore these aspects in our future work.

\section{Related Work}  \label{sec:related_works}

Despite the fact that malware developers and sophisticated adversaries are able to overcome air-gaps, most RTS were considered to be invulnerable against software security breaches; and  until recently security issues in RTS were not extensively discussed in industry or academia. The issues regarding information leakage through storage timing channels using architectural resources (\eg caches) shared between real-time tasks with different security levels is considered in the work \cite{sg1, sg2}. The authors proposed a modification to the fixed-priority scheduling algorithm and introduced a state cleanup mechanism to mitigate information leakage through shared resources. The cost paid for enforcing security mechanism is the reduced schedulability of overall system.

There has been some work \cite{xie2007improving, lin2009static}  on reconciling the addition of security mechanisms into real-time systems that considered periodic task scheduling where each task requires a security service whose overhead varies according to the quantifiable level of the service. A new scheduler \cite{xie2007improving} and enhancements to existing EDF scheduler \cite{lin2009static} is proposed to meet real-time requirements while maximizing the level of security achieved. In contrast, we consider a fixed-priority scheduling mechanism where security policies are executed sporadically using a server while meeting real-time requirements.

Although not in the context of security in RTS, a similar line of work to ours exists where the authors statically assign the  periods for multiple independent control tasks considering control delay as a cost metric  \cite{delay_period}. The cost functions are assumed to be linear in task periods. The control delay is estimated using an approximate response-time analysis and the authors presented an iterative procedure, where the actual (\eg nonlinear) cost functions are linearized around the current solution in each step. In contrast, our goal is to ensure security of the system without violating timing constraints of the real-time tasks. Hence, instead of minimizing response time, our goal is to assign best possible periods, so that the perturbation between  achievable period and desired period is minimized for all the security tasks.

The notion of randomization has been used in literature \cite{taskshuffler} with a view to hardening security mechanisms by minimizing predictability of deterministic RTS schedulers. The authors proposed a schedule obfuscation method that aimed at randomizing the task schedule while providing the necessary real-time guarantees for safe operation. It is not inconceivable that the randomization protocol in the work \cite{taskshuffler}  is complementary to those presented here and can be adopted to the proposed framework in order to make the system robust against attackers.  Different from our work at the scheduler level, architectural frameworks \eg \cite{securecore, slack_cornel} %for solving security problems such as intrusion detection 
aim to create hardware/software mechanisms to protect against security vulnerabilities. It is worth mentioning that these two sets of approaches could be combined to make the RTS more resilient to attacks.

%It is not inconceivable that the two sets of approaches could be combined to make the system more resilient to attacks.

\section{Conclusion} \label{sec:conclusion}

The evidence from recent successful attacks on automobiles \cite{ris_rts_1}, industrial control systems \cite{stuxnet} and  UAVs \cite{dronhack} indicate that RTS are not invulnerable to security breaches. 
%We presented methods for integrating security policies into real-time scheduling algorithms. 
In this work we are stepping towards developing an integrated security-aware RTS and   provide a glimpse of security design metrics for RTS. By using approaches such as the ones presented in this paper, designers of RTS are now able to improve their security posture and consequently safety -- which is the main goal for such systems.  This is also a step towards developing security metrics for the field of systems security in general.

\bibliographystyle{IEEEtran}
\bibliography{references}

\appendix

\subsection{Proof of Proposition \ref{prop:posy2}}  \label{appsec:proof_prop}

Let us rearrange %(\ref{eq:obj_server_nonlin}) 
Eq. (\ref{eq:ser_obj})
as $Q P ^{-1}$ which is clearly a posynomial. In order to reform the objective function 
%of $\mathbf{P}\mathbf{\ref{opt:server_non-lin}}$ 
in Eq. (\ref{eq:ser_obj}) as a standard GP minimization problem, we can rewrite 
Eq. (\ref{eq:ser_obj}) as
%(\ref{eq:obj_server_nonlin}) as 
\begin{equation}
\underset{Q, P}{\operatorname{min}}~  Q ^{-1} P
\end{equation}
which is also in posynomial form. Let us now rearrange %(\ref{eq:con1_server_nonlin}) 
Eq. (\ref{eq:ser_con1}) as follows
\begin{equation} \label{eq:posy_server1}
(Q + \Delta_{S}) P^{-1} \leq 1 
\end{equation}
where $\Delta_{S} =
  \sum\limits_{\tau_h \in hp(\tau_S)} (P + T_h) \cdot T_h^{-1} \cdot C_h $. Since the optimization variables (\eg capacity and replenishment period) are always positive, using the similar argument presented in Observation \ref{prop:posy1}, we can assert that Eq. (\ref{eq:posy_server1}) is a posynomial constraint.

\subsection{Proof of Theorem  \ref{th:posy}}  \label{appsec:proof_thm}
We prove the theorem using the geometric mean approximation \cite[Ch. 2]{gp_comm}. Since the denominator in Eq. (\ref{eq:server_bound_npoys}) is a posynomial, let us approximate $Q + T_i^*$ with a monomial by the following geometric mean approximation.

Let us denote $Q + T_i^*$ as $g(Q, T_i^*) = u_1(Q) + u_2(T_i^*)$ where $u_1(Q) = Q$ and $u_2(T_i^*) = T_i^*$. We can approximate $g(Q, T_i^*)$ with 
\begin{equation}
\hat{g}(Q, T_i^*) = {\left[ \frac{u_1(Q)}{a} \right]}^a \cdot {\left[ \frac{u_2(T_i^*)}{b} \right]}^b
\end{equation}
where $a = \frac{u_1(y_0)}{g(y_0, T_i^*)}$, $b = \frac{u_2(T_i^*)}{g(y_0, T_i^*)}$ and $y_0 \in \mathbb{R}^+$ is a constant that satisfies $\hat{g}(y_0, T_i^*) = g(y_0, T_i^*)$. The approximated monomial $\hat{g}(Q, T_i^*)$ can be rewritten as %follows
\begin{equation}
\hat{g}(Q, T_i^*) = {\left( \frac{Q}{a} \right)}^a \cdot {\left( \frac{T_i^*}{b} \right)}^b
\end{equation}
where $a = \frac{y_0}{y_0 + T_i^*}$, $b = \frac{T_i^*}{y_0 + T_i^*}$.
Using this monomial approximation, we can represent Eq. (\ref{eq:server_bound_npoys}) as 
\begin{equation}
\frac{P  (Q + I_i) + \Delta_{S}  Q}{Q \cdot \hat{g}(Q, T_i^*)} \leq 1 \quad \forall \tau_i \in \Gamma_S
\end{equation}
and the proof follows.

\end{document}